\documentclass[letter]{aa} 

\usepackage{graphicx}

\usepackage{caption}
\usepackage{subcaption}
\usepackage{siunitx}

\usepackage{txfonts}
\usepackage{color}

\begin{document}

\title{Insights on bar quenching from a multi-wavelength analysis: The case of Messier 95}



\author{K. George\inst{1}\fnmsep\thanks{koshyastro@gmail.com},  P. Joseph\inst{1,2}, C. Mondal\inst{2}, S. Subramanian\inst{2}, A. Subramaniam\inst{2}, K. T. Paul\inst{1}}


\institute{Department of Physics, Christ University, Bangalore, India \and Indian Institute of Astrophysics, Koramangala II Block, Bangalore, India}

  \abstract{ The physical processes related to the effect of bar in the quenching of star formation in the region between the nuclear/central sub-kpc region and the ends of the bar (bar-region) of  spiral galaxies is not fully understood. It is hypothesized that the bar can either stabilize the gas against collapse, inhibiting star formation or efficiently consume all the  available  gas,  with no  fuel  for  further  star  formation. We present a multi-wavelength study using the archival data of an early-type barred spiral galaxy, Messier 95, which shows signatures of suppressed star formation in the bar-region. Using the optical, ultraviolet, infrared, CO and HI imaging data we study the pattern of star formation progression, stellar/gas distribution and try to provide insights on the process responsible for the observed pattern. The FUV–NUV pixel colour map reveals a cavity devoid of UV flux in the bar-region that interestingly matches with the length of the bar ($\sim$ 4.2~kpc). The central nuclear region of the galaxy is showing a blue color clump and along the major-axis
  of the stellar bar the colour progressively becomes redder. Based on a comparison to single stellar population models, we show that the region of galaxy along the major-axis of the bar 
  (unlike the region outside the bar) is comprised of stellar populations with ages $\geq$ 350 Myr, with a star-forming clump in the center of younger ages ($\sim$ 150Myr). Interestingly the bar-region is also devoid of neutral and molecular hydrogen but with an abundant molecular hydrogen present at the nuclear region of the galaxy. Our results are consistent with a picture in which  the stellar bar in Messier 95 is redistributing the gas by funneling gas inflows to nuclear region, thus making the bar-region devoid of fuel for star formation.}
  

\titlerunning{Bar induced star formation quenching in M95}
\authorrunning{K. George\inst{1}}

\maketitle
%

\section{Introduction}

Galaxies in the local Universe follow a bimodal distribution in the optical broad band colors with the blue region mostly populated by star forming spiral galaxies and the red region dominated by elliptical/S0 galaxies with little or no ongoing star formation \citep{Strateva_2001,Baldry_2004}. However there exists a fraction of elliptical galaxies in the blue region \citep{Schawinski_2009} and spiral galaxies in the red region \citep{Masters_2010}. The number density of red galaxies are observed to increase from $z \sim 1$ which is now understood to be at the expense of blue galaxies \citep{Bell_2004,Faber_2007}. Several internal (AGN/stellar feedback, action of stellar bar) and external process (ram pressure stripping, major mergers, harassment, starvation, strangulation) have been proposed as responsible for the suppression of star formation (a process known as "quenching") that often involve morphological transformation of spiral galaxies (see \citealt{Peng_2015,Man_2018} and references therein). The existence of a population of passive red spiral galaxies (\citealt{vandenBerg_1976,Couch_1998,Dressler_1999,Poggianti_1999,Lee_2008,Cortese_Hughes_2009,Deng_2009,Masters_2010} and references therein) imply galaxies can transform from star forming to non star forming phase without invoking morphological transformation \citep{Fraser_2016}. Red spiral galaxies are found to have a higher optical bar fraction than blue spiral galaxies, which highlight the importance of stellar bars in quenching star formation \citep{Masters_2010,Masters_2011}. 

Stellar bars redistribute the disk content of galaxies via torques and can drive the secular evolution in spiral galaxies (\citealt{Combes_1981,Combes_1990,Debattista_2004,Kormendy_2004} and references therein). This is possible through the inflow of gas from the outer disk to the central region  which results  in an enhanced nuclear/central star formation observed in barred spiral galaxies (\citealt{Athanassoula_1992, Ho_1997,Sheth_2005,Coelho_2011,Ellison_2011,Oh_2012}). However, apart from the enhancement of star formation at the central regions the stellar bars can also suppress star formation (bar quenching) and is discussed in recent literature based on simulations as well as observations \citep{Masters_2010,Masters_2012,Cheung_2013,Gavazzi_2015,James_2016,Spinoso_2017,Khoperskov_2018,James_2018}. The stellar bar in massive star forming galaxies is understood to play a dominant role in regulating the red-shift evolution of specific star formation rates and mass dependent star formation quenching in field galaxies \citep{Gavazzi_2015}. The likelihood for disk galaxies hosting a bar is observed to be anti-correlated with specific star formation rate regardless of stellar mass and the prominence of bulge \citep{Cheung_2013}. Barred galaxies are also shown to have lower star formation activity relative to unbarred galaxies (\citealt{Consolandi_2017}, \citealt{Kim_2017}). They are found to be devoid of H$_{\alpha}$ flux in the radial range covered by the bar suggesting no ongoing or recent star formation \citep{James_2009}.\\

However the physical processes responsible for bar quenching are not well understood. There are primarily two mechanisms suggested for the quenching of star formation due to the effect of bars. During its formation the bar collects most of the gas inside the co-rotation radius. Then the bar induced shocks and shear can stabilize the gas against collapse by increasing turbulence and hence inhibit star formation (\citealt{Tubbs_1982}; \citealt{Reynaud_1998}; \citealt{Verley_2007}; \citealt{Haywood_2016}; \citealt{Khoperskov_2018}). Alternate mechanism is that, the bar induced torque drives gas inflows which enhance the nuclear star formation and making the region close to the bar devoid of fuel for further star formation (\citealt{Combes_1985}; \citealt{Spinoso_2017}). It is not certain which of these processes or a different unknown mechanism is responsible for star formation quenching in the region between the nuclear/central sub-kpc region and the ends of the bar (bar-region) of barred spiral galaxies. In the scenario of suppression of star formation by the stabilization of disk due to the bar induced torques, the gas from the bar-region of the galaxy need not be re-distributed/depleted. Thus the presence/absence of gas in the bar-region can put strong constraints on identifying the mechanism responsible for bar quenching in this galaxy. In this context here we present a multi-wavelength study based on the archival data of a barred spiral galaxy, Messier 95 (M95). \\

M95\footnote{$\alpha$(J2000) = 10:43:57.7 and $\delta$(J2000) = $+$11:42:14 according to Nasa/IPAC Extragalactic Database (NED).} (also known as NGC 3351) is a nearby (10 $\pm$ 0.4 Mpc, \citealt{Freedman_2001}) early-type barred spiral galaxy (Morphology; SBb). The angular scale of 1" corresponds to 48 pc at the distance of the galaxy.  M95 has stellar mass, HI mass,  H$_2$ mass and integrated star formation rate of $\sim$ 10$^{10.4}$ M$_{\odot}$, $\sim$ 10$^{9.2}$ M$_{\odot}$, $\sim$ 10$^{9}$ M$_{\odot}$ and $\sim$ 0.940 M$_{\odot}$/yr respectively \citep{Leroy_2008}. The gas phase metallicity (12 $+$ Log O/H) of M95 is 8.60 \citep{Ruyer_2014}. It is a nearly  face-on galaxy (inclination=41$\si{\degree}$, position angle=192$\si{\degree}$) with a prominent bar (See Figure~\ref{figure:fig1}). High quality multi-wavelength data of M95, ranging from radio to Ultraviolet (UV), are available. 
It shows nuclear star formation and hosts a star forming circumnuclear ring with a diameter of $\sim$ 0.7 kpc. This sub-kpc scale star formation is well studied in X-rays (\citealt{Swartz_2006}), UV (\citealt{Ma_2018}; \citealt{Colina_1997}), H$\alpha$ (\citealt{Planesas_1997,Bresolin_2002}) and near infrared (\citealt{Elmegreen_1997}). In a  multi-wavelength study, from Ultraviolet to mid-infrared, of the nuclear ring of M95, \citep{Ma_2018} presented the integrated properties of the ring and their correlation with bar strength. \cite{Mazzalay_2013} presented the properties of molecular gas within $\sim$ 300 pc of this galaxy using near infrared integral field spectrograph, SINFONI in Very Large Telescope and suggested that the nuclear region host a stellar population of a few Myr. H$\alpha$ imaging of larger area of M95 shows that the bar-region is devoid of emission (\citealt{James_2009}). The stellar population studies of this region indicate that they host an old population (\citealt{James_2016}). Long-slit spectroscopy of the bar-region showed a diffused emission which are not found to be associated with star formation. They attribute this emission to be due to post Asymptotic Giant Branch (p-AGB) stars (\citealt{James_2015}). These studies suggest that the observed nuclear star burst and  suppression of recent star formation ($\sim$ 10 Myr) in the bar-region is due to the effect of bar. However, the physical mechanism responsible for this observation is not understood. All the above points make this galaxy an excellent candidate to study the effect of bar on quenching star formation.
Throughout this paper, we adopt a flat Universe cosmology with $H_{\rm{o}} = 71\,\mathrm{km\,s^{-1}\,Mpc^{-1}}$, $\Omega_{\rm{M}} = 0.27$, $\Omega_{\Lambda} = 0.73$ \citep {Komatsu_2011}.\\

\section{Data and analysis}

\begin{figure}
\centering
\includegraphics[width=9cm,height=9cm,keepaspectratio]{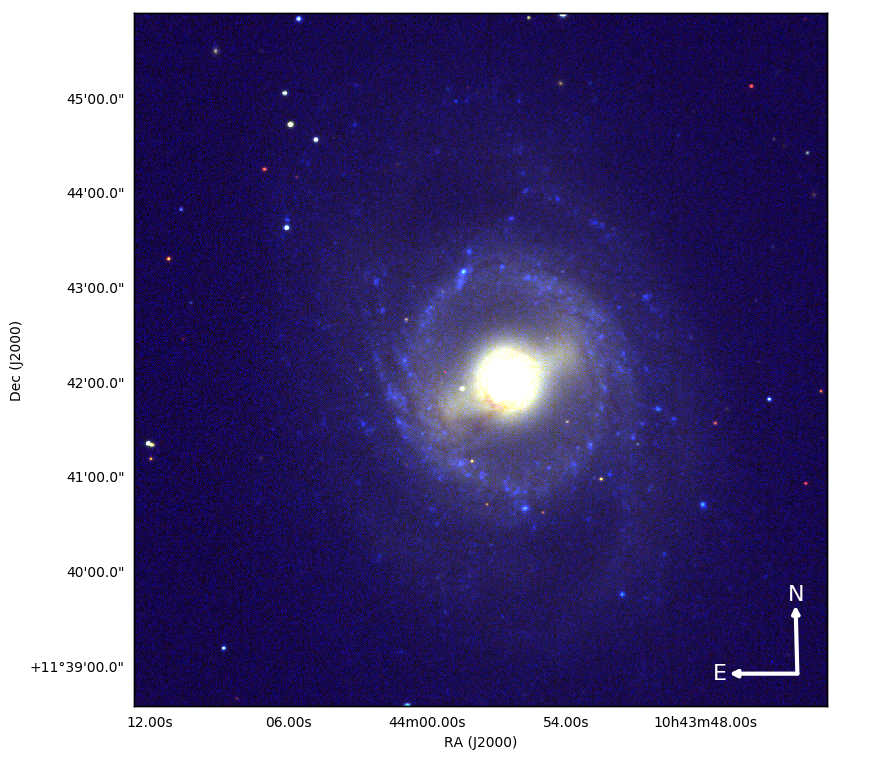}
\caption{Colour composite created from SDSS $urz$ filter pass band images of M95. The RGB image is created by assigning red ($z-$band), green ($r-$band) and blue ($u-$band) colours to the filter pass band images. The dust lane along the bar is seen in the color composite image.  
}\label{figure:fig1}
\end{figure}

In this study we exploit the archival data of M95 observed from ultraviolet to radio wavelengths as part of different campaigns.  We used the SDSS $urz$ DR9 (\citealt{Ahn_2012}) optical imaging data of M95 to construct a colour composite image which is shown in Figure~\ref{figure:fig1} and demonstrate the presence of a prominent stellar bar. The choice of blue ($u$) and red ($z$) pass band images helps in better visualizing the spatial variation of the relative contribution of young and evolved population of stars in the galaxy. The $u$ band flux is negligible in the bar-region, which instead is dominant in the region outside the stellar bar and could be hosting intense star formation which we address in detail using UV data.\\

M95 was observed in FUV ($\lambda_{eff}$=1538.6 {\AA}, Integration time=1692.2s) and NUV ($\lambda_{eff}$=2315.7 {\AA}, Integration time=1692.2s) wavelengths using the NASA GALEX mission  \citep{Martin_2005}. The GALEX FUV channel imaging is at $\sim$ 4.2" and the NUV channel imaging is at 5.3" resolution \citep{Morrissey_2007}. The FUV image is degraded to NUV resolution by running a Gaussian 2D kernel of width 0.57". The GALEX GR6/GR7 data of M95 field observed as part of Nearby Galaxy Survey (NGS) is pipeline reduced (with good photometric quality) and astrometry calibrated. We study the UV properties of this galaxy to probe recent star formation (past a few 100 Myr, \citealt{Kennicutt_2012}) over scales of $\sim$ 288~pc. The HI map of M95 from The HI Nearby Galaxy Survey (THINGS; \citealt{Walter_2008}) and the CO map (J$_{2-1}$ transition) from CO measured by HERA CO-Line Extragalactic Survey and Berkeley-Illinois-Maryland Association Survey of Nearby Galaxies (HERACLES; \citealt{Leroy_2009}) are used to understand the gas distribution. We used the infrared image from Spitzer IRAC 3.6 $\mu$ channel observed as part of S$^{4}$G \citep{Sheth_2010} to understand the distribution of evolved stellar population in the bar-region of the galaxy. \\

\begin{figure}
\centering
\includegraphics[width=9cm,height=9cm,keepaspectratio]{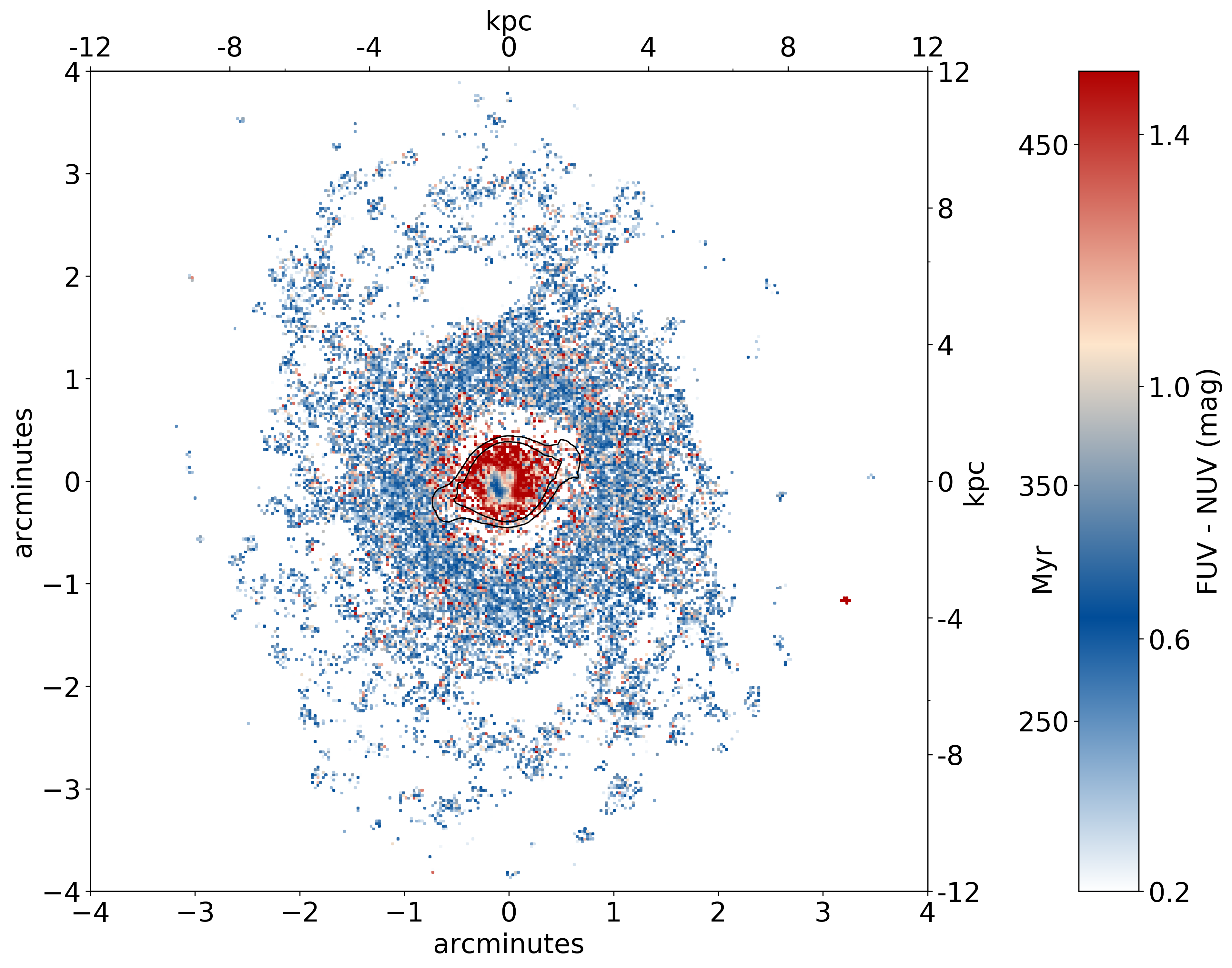}
\caption{$FUV-NUV$ colour map of the main body of M95. The pixels are colour-coded in units of $FUV-NUV$ colour. The corresponding single stellar population equivalent ages are also noted in the colour bar. The contour shows the stellar bar detected from the Spitzer IRAC 3.6 $\mu$ image of M95 with a length $\sim$ 4.2~kpc. The image measures $\sim$ 8$'$ $\times$ 8$'$ and corresponds to a physical size of $\sim$ 24~kpc on each side.}\label{figure:fig2}
\end{figure}

The foreground extinction from the Milky Way galaxy in the direction of M95 is $A_{V}$ = 0.076 \footnote{NED} \citep{Schlegel_1998} which we scaled to the FUV and NUV $\lambda_{mean}$ values using the \citet{cardelli_1989} extinction law and corrected the magnitudes.  The region of the FUV and NUV images that correspond to M95 was isolated using the threshold set by the background counts from the whole image. Pixels with values above the 3$\sigma$ of the threshold were selected to isolate the galaxy. The counts in the selected pixels were background subtracted, integration time weighted, and converted to magnitude units using the zeropoints of \citet{Morrissey_2007}. Magnitudes for each pixel are used to compute the FUV--NUV colour map of the galaxy (see Figure~\ref{figure:fig2}). The pixels are colour coded in units of FUV--NUV colour. The image is of size $\sim$ 8$'$ $\times$ 8$'$ and corresponds to a physical size of $\sim$ ~24kpc on each side at the rest-frame of the galaxy. The FUV--NUV colour map of M95 displays a redder region at the centre (with an embedded small blue clump) which is separated from the rest of the galaxy by a region with negligible UV flux. The redder region in Figure~\ref{figure:fig2} coincides with the major-axis of the bar of M95. It is interesting to note that the bar-region has negligible UV flux. This region also coincides with the region identified to be devoid of emission in $H\alpha$ \citep{James_2009}.\\

\begin{figure}
\centering
\includegraphics[width=7.5cm,height=7.5cm,keepaspectratio]{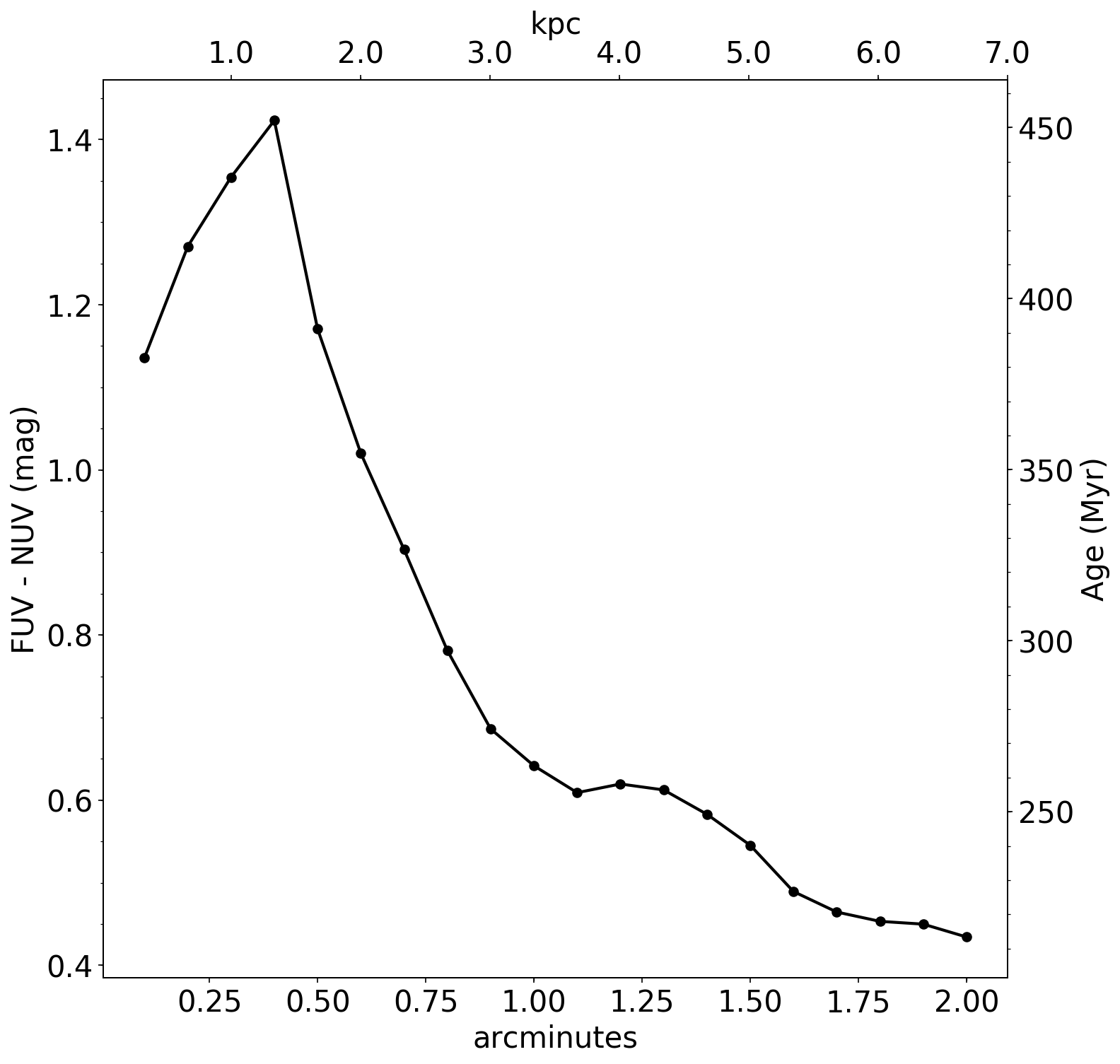}
\caption{Azimuthally averaged FUV--NUV colour profile of M95. The 2$'$ ($\sim$ 6 kpc) region of the galaxy been averaged in colour in concentric annuli of width 6". The profile shows an inner blue region gradually changing to redder colors, followed by a change to blue colours. The FUV--NUV colour values and the corresponding age estimates are shown on the left and right axes, respectively.}\label{figure:fig3}
\end{figure}

The  FUV--NUV colour map can be used to understand the star formation history of M95 and can, in particular, offer insights into the last burst of star formation. We used the {\tt Starburst99} stellar synthesis code to characterise the age of the underlying stellar population in M95 \citep{Leitherer_1999}. We selected 19 single stellar population (SSP) models over an age range of 1 to 900 Myr assuming a Kroupa IMF  \citep{Kroupa_2001} and solar metallicity (Z=0.02). The synthetic SED for a given age was then convolved with the effective area of the FUV and NUV pass-bands to compute the expected fluxes. The estimated values were then used to calculate the SSP ages corresponding to the observed FUV--NUV colours.  We performed a linear interpolation for the observed colour value and estimated the corresponding ages in all pixels in the FUV--NUV colour map. The ages for the FUV--NUV colour is shown in the colour bar in Figure~\ref{figure:fig2}. This exercise shows that the region along the major-axis of the bar hosts stellar populations of age $\geq$ 350 Myr and the nuclear/central sub-kpc region shows embedded bluer, younger clump of star formation ($\sim$ 150-250 Myr). Figure~\ref{figure:fig3} shows an azimuthally averaged colour profile of M95. The FUV--NUV colour has been measured in concentric annuli of width 6" ($\sim$ 0.3 kpc). We note the striking change in the colour profile moving outwards, where the colour change from blue to redder values in the very central region and finally to progressively bluer colours with increasing distance from the galaxy centre. There is a slight change to redder colors from 1'-to-1.5' away from the center of the galaxy. This is the region on the galaxy where the stellar bar meets the outer star forming region and is hosting dust lanes as seen in optical colour composite image (Figure~\ref{figure:fig1}). The FUV and NUV flux is subjected to extinction at the rest-frame of the galaxy. We do not have a proper extinction map of the galaxy M95. The FUV--NUV pixel colour maps and the derived ages can therefore be considered as the upper limits of the actual values.

\begin{figure}
\centering
\includegraphics[width=9cm,height=9cm,keepaspectratio]{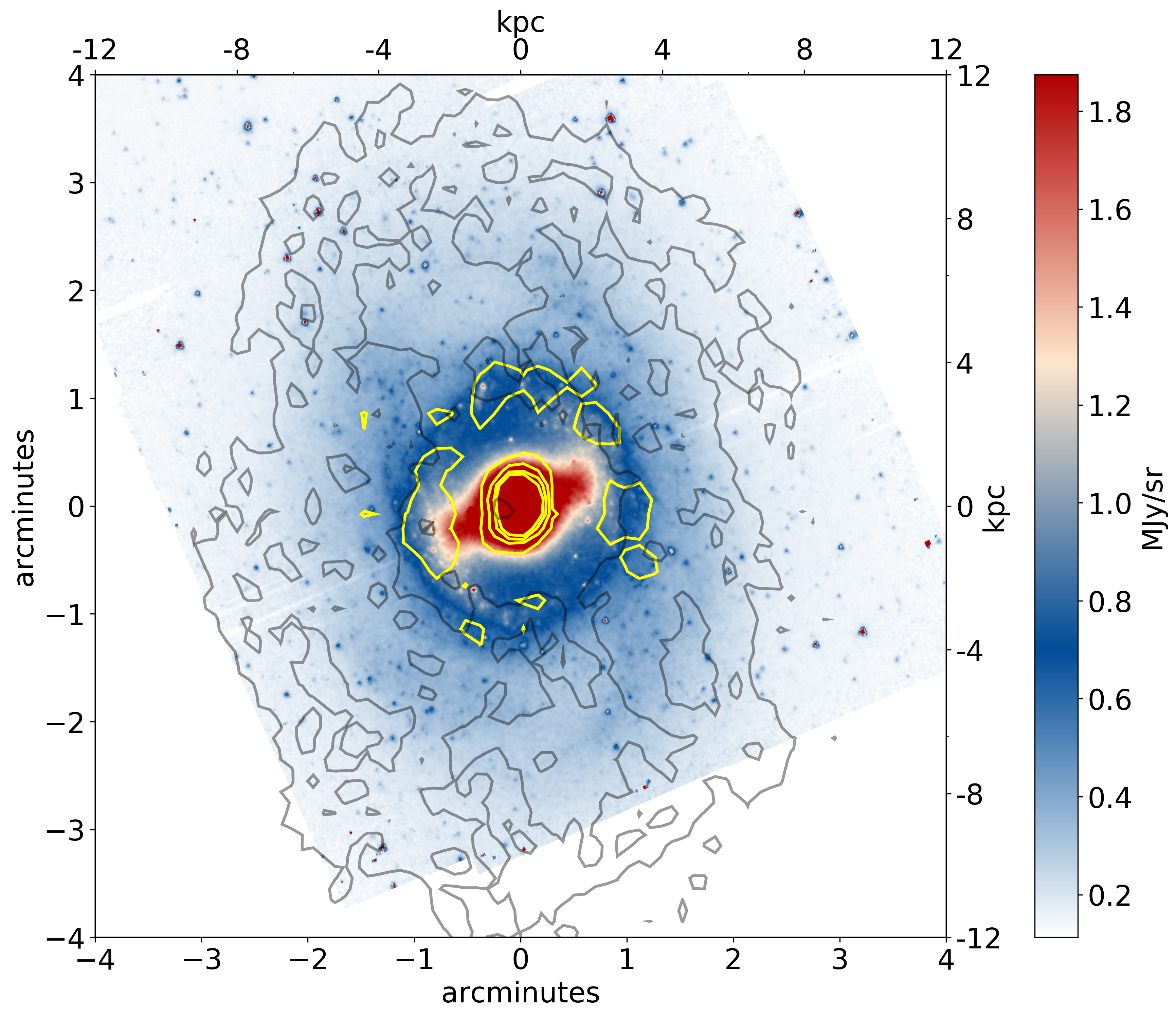}
\caption{The Spitzer IRAC 3.6 $\mu$ image (flux unit (MJy/sr)) of Messier 95 with the black contour demarcate stellar bar detected with a length $\sim$ 4.2~kpc. The HI contour (black colour) from THINGS (levels -0.66,20.22,41.10,61.99,82.87 in flux unit (Jy/beam/s)) and the CO contour (yellow colour) from HERACLES (levels -1.65, 2.27 in flux unit (K Km/s)) are overlaid.  There is an offset between the HI and CO emission peak at the center of the galaxy. The color scale is adjusted such that the stellar bar feature is prominently seen in the IRAC 3.6 ${\mu}$ image.}\label{figure:fig4}
\end{figure}

The Spitzer IRAC 3.6 ${\mu}$ image of a galaxy can be used as a extinction free tracer for the evolved stellar population which dominate the underlying stellar mass \citep{Meidt_2014}. The Spitzer IRAC 3.6 ${\mu}$ image of M95 is shown in Figure~\ref{figure:fig4} with appropriate scaling to enhance the appearance of the stellar bar. We note that the stellar bar is prominent in infrared image and could be hosting evolved stellar population. The length of the stellar bar from the infrared image is $\sim$ 87" ($\sim$ 4.2 Kpc). The HI contours (black colour) and the CO contours (yellow colour) are overlaid over the Spitzer image. Comparing Figure~\ref{figure:fig2} and Figure~\ref{figure:fig4} it is interesting to see that the 4.2~kpc diameter circular region (interestingly the length covered by stellar bar), avoiding the central nuclear region, lacks molecular/neutral hydrogen and star formation. The central sub-kpc nuclear region of the galaxy is hosting significant molecular gas content, star formation and some amount of neutral hydrogen.\\

\section{Discussion}

The stellar bar can channel the gas inwards of the central regions of the galaxy within which star formation can happen and is proposed to be responsible for the formation of pseudo bulge \citep{Sanders_1976,Roberts_1979,Athanassoula_1992,Ho_1997,Kormendy_2004, Jogee_2005, Lin_2017, Spinoso_2017}. On the other hand the bar can also suppress recent star formation in galaxy disks \citep{James_2016,Spinoso_2017,James_2018}. The recent simulations demonstrate that the stellar bar to be efficient in quenching star formation with a reduction in star-formation rate by a factor of ten in less than 1 Gyr \citep{Khoperskov_2018}. Simulations also predict stellar bars as long lived features in isolated disc galaxies with life time up to $\sim$ 1000 Myr \citep{Athanassoula_2013}. This imply that stellar bars can keep the galaxy quenched for at least $10{^9}$ yr and could be a dominant mechanism in shutting down star formation in galaxies over all redshift. It is therefore necessary to make a detailed understanding of the processes operating during the bar quenching in galaxies. 

There are primarily two mechanisms suggested for the quenching of star formation due to the effect of bars. The stellar bar in the galaxy can either stabilize the disk against collapse, inhibiting star formation (\citealt{Tubbs_1982,Reynaud_1998,Verley_2007,Haywood_2016,Khoperskov_2018}), or efficiently consume all the available gas, with no fuel for further star formation (\citealt{Combes_1985,Spinoso_2017} ). The turbulence set by the bar prevents the fragmentation of molecular gas within the co-rotation radius and thus suppresses star formation in the bar-region of the galaxy. In such a scenario the gas in the bar-region of the galaxy need not be depleted nor re-distributed to quench star formation. The presence or alternatively the absence of the neutral and molecular hydrogen in the quenched barred galaxies can provide insights regarding the mechanisms responsible for bar quenching. We note that in the scenario where  the suppression of star formation due to bar induced turbulence, it is not clear whether all the gas will be shock heated. Signatures of shock heating should be seen in H$\alpha$ observations.\\

The multi-wavelength study of M95 based on the archival data ranging from ultraviolet, optical, infrared, neutral hydrogen and molecular hydrogen, as traced by CO, paint a picture of star formation quenching happening in the bar-region. There is no star formation in the last 100-200 Myr as evident from the FUV--NUV colour map. The lack of molecular and neutral hydrogen in this region implies that the stellar bar might have re-distributed the gas. The stellar bar can funnel the gas to the  center and can be the reason for significant molecular gas content and recent star formation observed in the central sub-kpc nucelar/central region. This can lead to  nuclear star bursts and formation of substructures (such as circumnuclear rings). M95 is known to have such features (\citealt{Colina_1997};\citealt{Ma_2018}). We note here that the barred galaxies are demonstrated to have an enhanced star formation at the center \citep{Ellison_2011} and in the case of M95 also it is observed to have younger age clumps ($\sim$ 150-250 Myr). 
This funneling of gas to the central sub-kpc region would have depleted gas in the bar-region. Hence suppressed star formation due to lack of fuel. On the other hand, there is significant neutral hydrogen present outside the length of bar along with presence of young stellar population. \\

The absence of CO and HI in the bar-region of M95 can be considered as a support to the scenario of gas re-distribution. The scenario of gas heating due to the stabilization of disk by bar induced torques can prevent gas cooling which in turn can inhibit star formation. However we expect to see the signature of such a gas heating in the form of  significant H$\alpha$ emission which is lacking along the stellar bar (within the detection limits) in the case of M95 as demonstrated by the H$\alpha$ imaging observations of \citet{James_2009} (see also \citealt{James_2015} where a diffuse emission in H$\alpha$ and [NII] 6584 $\AA$ is attributed to p-AGB stars but shocks are not completely ruled out).\\

We present here evidence for gas re-distribution due to the stellar bar and subsequent star formation quenching within the bar co-rotation radius in M95. The main result of our analysis is a region, between the nuclear region and the ends of the bar, devoid of gas and star formation in the past a few 100 Myr. Star formation is quenched in this region and the absence of  molecular/neutral hydrogen gas imply no further star formation is possible or in other words bar quenching is a dominant star formation suppression mechanism in M95. In the absence of an external supply of gas, the star formation in the center will deplete the molecular hydrogen completely and the galaxy will be eventually devoid of star formation in the bar and the central nuclear region.

It is not clear whether bar quenching is the dominant process responsible for star formation suppression in barred spiral galaxies in general and the redistribution of the gas due to stellar bar the main governing process. The pilot study reported here demonstrate the capability of multi-wavelength analysis in understanding the role of stellar bar in star formation progression and gas distribution in spiral galaxies. The results presented here calls for a detailed analysis of a statistically large sample of face-on barred galaxies with multi-wavelength observations which will be reported in a forthcoming paper. The stellar and gaseous kinematics (ionized gas) along the region of bar can be understood in more detail from the observations based on ongoing optical IFU surveys.\\

\section{Summary}

We present here observational evidence for star formation quenching due to the presence of a stellar bar and the mechanism responsible for quenching in galaxy Messier 95 based on a multi-wavelength analysis using the archival data. Based on the FUV--NUV pixel colour map we demonstrate that the central 4.2 kpc diameter region along the stellar bar of galaxy is composed of stellar population with equivalent ages $\geq$ 350 Myr. This imply that currently there is no ongoing star formation along the region covered by the bar. The central sub-kpc region of the galaxy is hosting abundant supply of molecular hydrogen with the region along the bar devoid of neutral and molecular hydrogen, but is present outside the stellar bar-region. This is a direct evidence coming from observations for the stellar bar in Messier 95 dynamically re-distributing the gas making the region close to the bar devoid of fuel for star formation. A similar analysis along with a spatially resolved study of the gaseous and stellar kinematics on a statistically significant number of barred galaxies can give more insights on bar quenching in spiral galaxies.

\begin{acknowledgements}
We thank the anonymous referee for the comments, which improved
the scientific content of the paper. KG acknowledge the stimulating discussion within the GASP collaboration on stellar bars in galaxies. This work made use of THINGS, 'The HI Nearby Galaxy Survey' \citep{Walter_2008}. This work made use of HERACLES, `The HERA CO-Line Extragalactic Survey’ \citep{Leroy_2009}. This research made use of Astropy, a community-developed core Python package for Astronomy \citep{Astropy_Collaboration_2018}. Funding for SDSS-III has been provided by the Alfred P. Sloan Foundation, the Participating Institutions, the National Science Foundation, and the U.S. Department of Energy Office of Science. The SDSS-III web site is http://www.sdss3.org/. SDSS-III is managed by the Astrophysical Research Consortium for the Participating Institutions of the SDSS-III Collaboration including the University of Arizona, the Brazilian Participation Group, Brookhaven National Laboratory, Carnegie Mellon University, University of Florida, the French Participation Group, the German Participation Group, Harvard University, the Instituto de Astrofisica de Canarias, the Michigan State/Notre Dame/JINA Participation Group, Johns Hopkins University, Lawrence Berkeley National Laboratory, Max Planck Institute for Astrophysics, Max Planck Institute for Extraterrestrial Physics, New Mexico State University, New York University, Ohio State University, Pennsylvania State University, University of Portsmouth, Princeton University, the Spanish Participation Group, University of Tokyo, University of Utah, Vanderbilt University, University of Virginia, University of Washington, and Yale University.
\end{acknowledgements}


\end{document}